\documentclass{article}
\usepackage[utf8]{inputenc}
\usepackage[english]{babel}

\usepackage[letterpaper,top=2cm,bottom=2cm,left=3cm,right=3cm,marginparwidth=1.75cm]{geometry}

\usepackage{textgreek}

\usepackage{graphicx}%
\usepackage{multirow}%
\usepackage{rotating}
\usepackage{amsmath,amssymb,amsfonts}%
\usepackage{amsthm}%
\usepackage{mathrsfs}%
\usepackage[title]{appendix}%
\usepackage{xcolor}%
\usepackage{threeparttable}
\usepackage{textcomp}%
\usepackage{manyfoot}%
\usepackage{booktabs}%
\usepackage{algorithm}%
\usepackage{algorithmicx}%
\usepackage{algpseudocode}%
\usepackage{listings}%
\usepackage{subcaption}
\usepackage{float}
\usepackage{abstract}
\usepackage{url}

\usepackage[colorlinks=true, allcolors=blue]{hyperref}
\usepackage{silence}
\providecommand{\keywordsname}{Keywords} 
\newcommand{\keywords}[1]{\textbf{\keywordsname: }#1}

\title{Depressive symptoms are reflected differently across digital contexts}
\author{Yajing Wang, Emilia Marchese, Talayeh Aledavood, Juhi Kulshrestha}
\date{}
\begin{document}
\maketitle

\begin{abstract}
As more of everyday life takes place online, digital behavior may provide a potential window into how depressive symptoms are reflected in daily life. Yet digital mental health studies have produced mixed findings. These inconsistencies may partly reflect how digital behavior is measured: self-reported use, single-device studies, and aggregate screen time can obscure differences across devices, activities, and patterns of engagement. We combined monthly assessments of depressive symptoms with passively recorded mobile and desktop web traces from 1,146 adults in Germany over six months. We examined how general, cognitive–affective, and somatic depressive symptoms are reflected across digital contexts defined by device and activity type. Associations varied markedly across these contexts. On mobile, more severe symptoms were associated with more nighttime activity, greater use of social media, messaging, and entertainment, and fewer but longer sessions. On desktop, associations were fewer and largely involved reduced engagement with news, shopping, and adult content. Mobile associations primarily arose for general and cognitive–affective symptoms, whereas desktop associations were concentrated in somatic symptoms. 
Our findings suggest that characterizing how depressive symptoms are reflected in digital behavior requires attending to what people do online and where, not only how much screens are used.
\end{abstract}

\keywords{Depressive symptoms, Social Media, Web trace, Online behavior, Digital well-being}

\section{Introduction}

Depression is a debilitating disorder that affects approximately $5.7\%$ of adults worldwide and substantially increases suicide risk~\cite{who_depression_2025}.
It fundamentally alters how people interact with everyday activities, characterized by reduced engagement and increased avoidance~\cite{apa2022dsm5tr,WHO2024CDDR,VolovicShushan2026EMA}. 
Increasingly, 
daily life unfolds online, where social, professional, recreational, 
and informational activities have migrated from physical to digital 
settings~\cite{harvey2022screenbased,comparitech2025screen}. 
Traditional diagnostic frameworks, however, remain 
disconnected from this shift. While effective for overall assessment of 
depression~\cite {apa2022dsm5tr,WHO2024CDDR}, they generally do not incorporate questions about individuals' behaviors on digital world. 
This distinction matters because online and offline 
interactions entail distinct behavioral and emotional demands
~\cite{trand2020peer,lieberman2020twosocial,adwi2025onlineoffline}.
Consequently, how depression manifests in digital behavior represents a critical, underexplored frontier. Here, we investigate how depression manifests in individuals' digital behaviors and explore whether these behaviors offer a complementary window into depression in everyday life.

While researchers increasingly examine digital behavioral markers of depression, existing evidence remains mixed. Some studies link depression to greater phone and social media use~\cite{saeb_mobile_2015,ivie_meta-analysis_2020,boers_association_2019}, while others connect it to reduced digital engagement, such as fewer applications and shorter session durations~\cite{yang_association_2023,leaning2024smartphone}, or find limited association between online behavior and mental well-being~\cite{huang2017time,shin2022online,winbush2025smartphone}.
Much of this inconsistency may stem from three main limitations in how digital behavior has been captured and measured. 
First, many studies rely on self-reported measures of digital technology use~\cite{ivie_meta-analysis_2020,boers_association_2019,bettmann2021young}, a metric that often diverges from objective tracking data~\cite{parry_systematic_2021}. For example, users systematically misjudge their daily screen time and broad digital media consumption~\cite{araujo_how_2017}
Second, focusing on a single device, typically the mobile phone~\cite{leaning2024smartphone,saeb_mobile_2015}, can overlook desktop use and treat one device as representative of broader digital behavior. This matters because desktop and mobile environments are complementary rather than interchangeable, supporting different tasks, contexts, purposes, and interaction patterns~\cite{Jokela2015device,karlson2009working,Oulasvirta2012Habits,Church2011MobileWeb}. 
Third, studies often aggregate all activities into a single ``screen time'' metric, erasing distinct purposes and experiences~\cite{saeb_mobile_2015}, or focus exclusively on specific categories like social media~\cite{bettmann2021young,yoon_is_2019}, ignoring the broader digital ecosystem. 
These measurement limitations may contribute to inconsistent findings and obscure an important question: whether depression is associated not simply with more or less technology use, but with different engagement patterns across digital contexts.

To overcome these gaps, our study uses fine-grained, passively-collected web trace data to capture a more accurate picture of how depressive symptoms are reflected in various digital contexts, which are defined as combinations of device and activity type.
First, we replace self-reported web use metrics with continuous, passively logged web and application activity, eliminating inaccuracies from subjective recall~\cite{parry_systematic_2021,araujo_how_2017}. 
Second, we analyze mobile and desktop usage separately rather than treating a single device as representative of unified digital behavior. 
Finally, we categorize usage across a wide range of web activity types rather than isolating specific categories or collapsing all activity into a single ``screen time'' metric. 
This approach allows us to examine whether and how behavioral changes associated with depressive symptoms surface in everyday web behavior across these digital contexts.

We combined continuous, objective, passively-collected web trace data from $1,146$ adults across mobile and desktop devices over six months (782 with mobile, 636 with desktop) with six monthly waves of the Patient Health Questionnaire (PHQ-9) ~\cite{kroenke2001phq}, a validated measure of depression severity commonly used for screening and monitoring in clinical practice. 
Analyzing each device environment separately, we used linear mixed-effects models (LMMs) to relate individuals' web use behavior to their depression severity and specific symptom dimensions at the population level, controlling for gender, age, income, chronotype, and survey wave. 
We further compared individuals with persistently high versus persistently minimal symptoms. 
Additionally, we examined whether these associations also hold within individuals over time.

Our analysis suggests that depression is not associated with a uniform change in how much people spend time online, but with decreases and increases that depend on the environment. 
In the desktop environment, higher depression is associated with spending less time on activities such as shopping and news. 
In the mobile environment, associations are positive, with higher depression linked to more time on entertainment, messaging, and social media. 
More broadly, these findings suggest that the useful question is not how much time people spend on screens, but what the behavioral profile of depression looks like within each digital context. 
Because that profile differs across environments, digital indicators of depression need to be environment-specific rather than assumed to represent digital behavior as a unified whole or to generalize from one environment to another.

\section{Study design and measurement}
\begin{figure*}[t!]
\centering
\includegraphics[width=\textwidth]{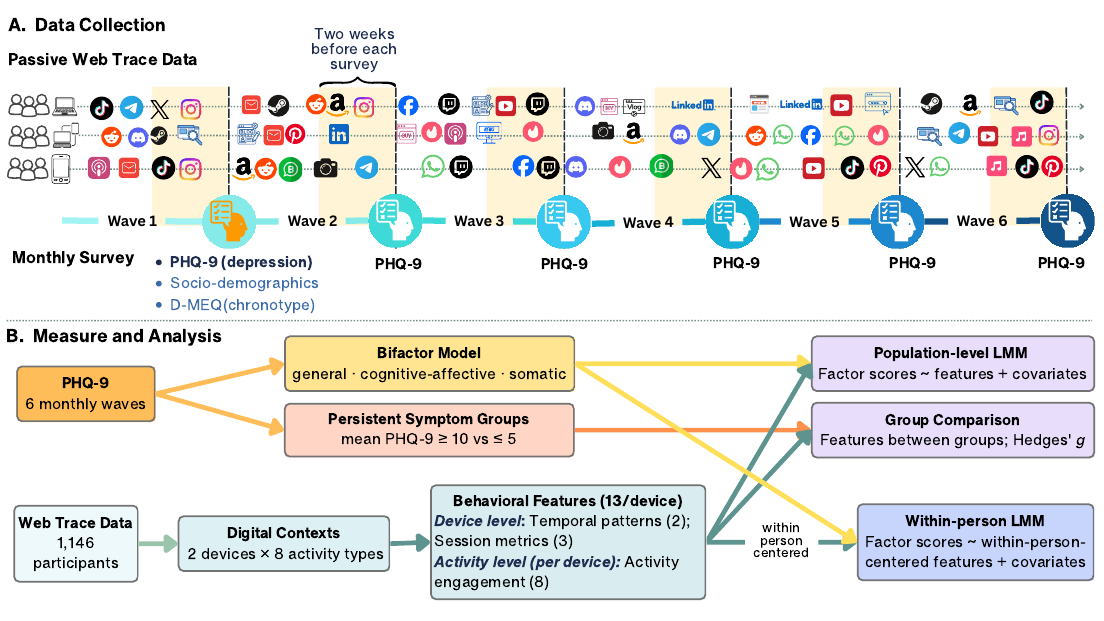}
\caption{\textbf{Study design.} (\textit{A}) Data collection: $1,434$ German adults completed PHQ-9 at six monthly waves while their mobile and desktop web browsing was continuously recorded over six months; the two-week window before each survey (shaded) was used to compute behavioral features. (\textit{B}) Analysis pipeline: PHQ-9 responses were scored with a bifactor model yielding general, cognitive--affective, and somatic factors, and web traces were used to derive behavioral features. Both measures were analyzed using population-level LMMs, group comparisons, and exploratory within-person analyses, separately for each device.}
\label{fig:design}
\end{figure*}

This study used a six-month longitudinal panel with web trace data that we collected passively, combined with repeated depression assessments among German adults ( Figure~\ref{fig:design}). 
We had $1,146$ participants after quality control, comprising $782$ with mobile web trace data and $636$ with desktop web trace data (see \textit{Materials and Methods} for exclusion criteria).
Participants completed the PHQ-9 once per month for six consecutive waves, with each survey response aligned to online activity recorded during the preceding two weeks to match the PHQ-9 reference period.

We used the PHQ-9 responses in two complementary ways. 
First, for group comparisons, participants were classified into persistent symptom groups using established clinical cutoffs~\cite{kroenke2001phq}: those with a mean PHQ-9 score of $10$ or higher across waves were assigned to the moderate-to-severe symptom group, and those with a mean score of $5$ or lower were assigned to the minimal-symptom group. 
Second, we modeled the PHQ-9 symptom structure using a bifactor model that yielded continuous scores for a general depression factor together with two specific symptom dimensions: cognitive--affective symptoms (anhedonia, depressed mood, worthlessness, concentration difficulties, and suicidal ideation) and somatic symptoms (sleep disturbance, fatigue, appetite change, and psychomotor change) (see \textit{Materials and Methods} for model specification and validation). These factor scores were used in longitudinal analyses to examine how online activities were associated with overall depression and its symptom dimensions.

Digital behavior was measured using passively collected mobile and desktop usage logs, recording application/web domain, visit timestamp, and its duration. 
We processed mobile and desktop data separately, as these devices support distinct forms of online engagement, rendering them distinct digital environments rather than interchangeable proxies~\cite{Jokela2015device,karlson2009working}. 
To move beyond aggregate usage metrics, we classified observed domains and applications into eight functional categories: social media, messaging, entertainment, news, shopping, productivity, adult content, and gaming (classification details in \textit{Materials and Methods}).

Per device and assessment wave, we derived 13 behavioral features from the two weeks preceding PHQ-9 administration. 
These features measured how each person behaves within their digital contexts in three facets: activity engagement, session metrics, and temporal patterns. 
Eight category-specific features measured daily active time for each functional category. 
Three session metrics quantified activity distribution and session structure: mean daily session count, mean session duration, and mean inter-session interval.
Two temporal features captured activity timing: nighttime activity ratio (00:00–06:00) and interdaily stability (consistency of timing across days; definitions in \textit{Materials and Methods}).

We examined the depression–behavior relationship with two complementary analytic approaches, with mobile and desktop behaviors analyzed separately throughout. 
First, we fit LMMs to examine associations between behavioral features and three depression factors (general depression, cognitive–affective symptoms, and somatic symptoms), controlling for gender, age, income, chronotype, and survey wave. 
Second, we compared individuals in persistent moderate-to-severe versus minimal-symptom groups to identify behavioral differences between consistently elevated and consistently minimal depressive symptom profiles. 
As a complementary within-person check, we tested whether associations persisted within individuals over time (i.e., as deviations from each participant's own average across survey waves). 
This design provides a framework for examining how patterns of web engagement relate to depressive symptoms across digital contexts, both continuously across the population and between individuals with persistently different symptom levels (complete analytical details in \textit{Materials and Methods}).

\section{Results}

To examine whether depression is associated with patterns of web use, specifically activity type and device context, we analyzed mobile and desktop data separately.

\subsection{Higher depression is associated with later, more social- and entertainment-focused mobile use}
For mobile devices, we modeled the relationship between three depression symptom dimensions (general, cognitive--affective, and somatic) and application/web use using LMMs and group comparisons contrasting participants with persistent moderate-to-severe (mean PHQ-9 $\geq 10$; $n = 177$) versus minimal symptoms (mean PHQ-9 $\leq 5$; $n = 392$). 
Each LMM modeled a depression factor score as a function of behavioral features, controlling for age, gender, income, chronotype, and survey wave (random intercepts for participants; see \textit{Linear Mixed-Effects Models} for detailed specifications). 
Behavioral features accounted for modest variance in symptom scores (marginal $R^2 = 0.03$--$0.10$, conditional $R^2 = 0.47$--$0.76$; see \textit{SI Appendix}, Table S9).

LMM results (Table~\ref{tab:lmm}, top) indicated that fewer daily mobile sessions were associated with higher general depression ($\beta = -0.020$, 95\% CI $[-0.039, -0.002]$, $p = 0.032$). 
However, group comparisons showed that reduced session frequency reflected longer continuous engagement: the moderate-to-severe symptom group exhibited significantly longer average session duration than the minimal-symptom group (0.91 versus 0.70 hr; Hedges' $g = 0.372$, 95\% CI $[0.190, 0.553]$, $p_{\text{adj}} = 0.002$). 
This prolonged engagement also occurred disproportionately at night: a larger proportion of activity during nighttime was associated with higher general depression scores ($\beta = 0.282$, 95\% CI $[0.079, 0.484]$, $p = 0.007$). 
Group comparisons confirmed this temporal shift (Fig.~\ref{fig:forest}), with the moderate-to-severe group having a larger nighttime activity ratio (0.15 versus 0.11; $g = 0.335$, 95\% CI $[0.148, 0.521]$, $p_{\text{adj}} = 0.005$).

Both approaches revealed consistent activity-pattern findings. 
The moderate-to-severe group exhibited higher daily engagement with social media (0.57 versus 0.42 hr/day; Hedges' $g = 0.230$, 95\% CI $[0.047, 0.416]$, $p_{\text{adj}} = 0.045$), messaging (0.37 versus 0.27 hr/day; $g = 0.240$, 95\% CI $[0.046, 0.435]$, $p_{\text{adj}} = 0.045$), and entertainment platforms (0.37 versus 0.21 hr/day; $g = 0.288$, 95\% CI $[0.095, 0.478]$, $p_{\text{adj}} = 0.023$). 
Correspondingly, greater use of entertainment applications was associated with higher cognitive--affective symptom scores ($\beta = 0.046$, 95\% CI $[0.007, 0.084]$, $p = 0.022$).

In an exploratory within-person analysis, similar associations appeared over time.
When participants exceeded their typical mobile entertainment use, they reported higher cognitive--affective symptom scores ($\beta = 0.054$, 95\% CI $[0.001, 0.108]$, $p = 0.048$). 
Similarly, elevated nighttime activity share ($\beta = 0.232$, 95\% CI $[0.018, 0.446]$, $p = 0.034$) and above-typical adult content consumption ($\beta = 0.738$, 95\% CI $[0.127, 1.345]$, $p = 0.018$) were each associated with higher general symptom scores.
These within-person associations converge with the between-person results, suggesting that depression-related shifts in digital behavior reflect both stable individual differences and dynamic within-individual variation.

\subsection{Higher depression is associated with reduced news, shopping, and adult content use on desktop} 
For desktop, we applied the same analytic framework using LMMs and group comparisons (moderate-to-severe: $n = 125$; minimal-symptom: $n = 337$). 
In contrast to mobile, desktop analyses yielded fewer significant associations, predominantly in the inverse direction. 
At the population level, the associations were only with somatic symptoms. 
As on mobile, behavioral features explained modest variance (marginal $R^2 = 0.03$--$0.10$, conditional $R^2 = 0.42$--$0.75$; see \textit{SI Appendix}, Table S9). 
This divergence aligns with evidence that desktop use involves more task-oriented rather than pleasure-oriented activities compared to mobile. 
Lower engagement with desktop shopping ($\beta = -0.088$, 95\% CI $[-0.170, -0.006]$, $p = 0.033$), news reading ($\beta = -0.121$, 95\% CI $[-0.241, -0.001]$, $p = 0.049$), and adult content consumption ($\beta = -0.089$, 95\% CI $[-0.142, -0.036]$, $p = 0.001$) was associated with higher somatic symptoms, whereas greater entertainment use was the only exception, showing a positive association ($\beta = 0.041$, 95\% CI $[0.007, 0.075]$, $p = 0.017$). 
General and cognitive--affective symptoms showed no population-level associations with desktop activity (Table~\ref{tab:lmm}).

\begin{figure}[H]
\centering
\includegraphics[width=\linewidth]{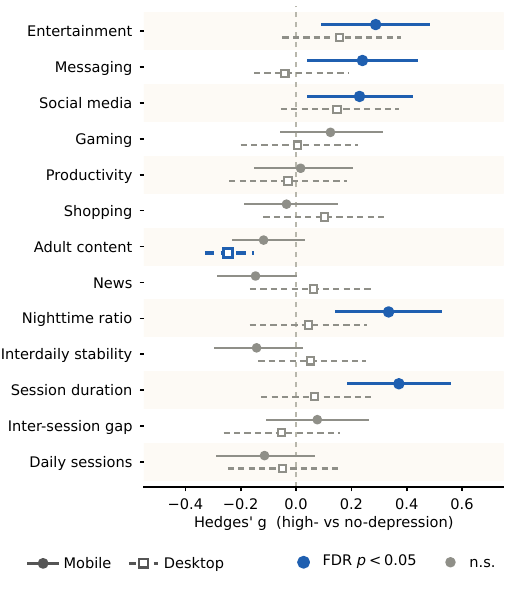}
\caption{\textbf{Group differences in web behavior between moderate-to-severe (mean PHQ-9 $\geq 10$) and minimal-symptom (mean PHQ-9 $\leq 5$) groups.} Mobile: filled circles, solid lines; desktop: open squares, dashed lines. Points show Hedges' $g$ with 95\% CI; blue indicates FDR-significant differences ($p_{\text{adj}} < 0.05$), gray non-significant. Positive values denote greater use in the moderate-to-severe group. (see \textit{SI Appendix}, Table S11 for full results)}

\label{fig:forest}
\end{figure}

Group comparisons confirmed this pattern (Fig.~\ref{fig:forest}), with groups differing only in adult content consumption: the moderate-to-severe group spent less time than the minimal-symptom group (0.01 versus 0.08 hr/day; Hedges' $g = -0.245$, 95\% CI $[-0.329, -0.150]$, $p_{\text{adj}} = 0.006$). 
Within-person analyses replicated this pattern across waves. 

When participants spent less time than usual on productive activities, they reported higher cognitive--affective symptoms ($\beta = -0.141$, 95\% CI $[-0.269, -0.013]$, $p = 0.032$); less time than usual on shopping ($\beta = -0.104$, 95\% CI $[-0.202, -0.006]$, $p = 0.037$) or adult content ($\beta = -0.100$, 95\% CI $[-0.162, -0.038]$, $p = 0.002$) corresponded with higher somatic symptoms; and more entertainment time was associated with slightly higher somatic symptoms ($\beta = 0.048$, 95\% CI $[0.006, 0.093]$, $p = 0.026$).

\begin{table}[t!]
\centering
\small
\setlength{\tabcolsep}{4pt}

\begin{tabular}{llrcr}
\toprule
Factor & Feature & $\beta$ & 95\% CI & $p$ \\
\midrule
\multicolumn{5}{l}{\textbf{Mobile}} \\
General & Nighttime usage ratio & 0.282 & [0.079, 0.484] & 0.007 \\
General & Avg.\ daily sessions (n) & $-$0.020 & [$-$0.039, $-$0.002] & 0.032 \\
Cog--affective & Entertainment & 0.046 & [0.007, 0.084] & 0.022 \\
\midrule
\multicolumn{5}{l}{\textbf{Desktop}} \\
Somatic & News & $-$0.121 & [$-$0.241, $-$0.001] & 0.049 \\
Somatic & Shopping & $-$0.088 & [$-$0.170, $-$0.006] & 0.033 \\
Somatic & Adult content & $-$0.089 & [$-$0.142, $-$0.036] & 0.001 \\
Somatic & Entertainment & 0.041 & [0.007, 0.075] & 0.017 \\
\bottomrule
\end{tabular}
\caption{\textbf{Population-level associations between web behavior and 
depression factors (linear mixed-effects models).} Each row represents one feature--factor association. 
$\beta > 0$ ($\beta < 0$): greater (reduced) behavior with higher symptoms. Only significant 
associations are shown; models adjust for age, gender, income, 
chronotype, and wave (see \textit{SI Appendix}, Table S7, S8 for full results).}

\label{tab:lmm}
\end{table}

\subsection*{Depression-related web behavior patterns differ across 
environments and symptom dimensions}

Mobile and desktop environments exhibited different patterns rather than uniform changes in total usage volume. 
On mobile, higher depression correlated with greater nighttime activity and increased time on social media, messaging, and entertainment platforms. 
On desktop, associations were fewer and reflected reduced engagement with news, shopping, and adult content. 
The three symptom dimensions varied across environments differently. 
Mobile web use primarily related to general depression (in terms of usage volume and temporal patterns) and cognitive--affective symptoms (in activity types, particularly entertainment). 
Desktop web use was predominantly associated with somatic symptoms across multiple activities (news, shopping, adult content, and entertainment).
Overall, depression was linked more to activity patterns and context than total 
usage, varying systematically across symptom dimensions.

\section*{Discussion}

By continuously and objectively recording the mobile and desktop web traces of 1,146 adults across six months, we found that overall depression severity and its specific dimensions were not linked to a uniform rise or decrease in digital use; rather, the direction of the association depended on the digital context. 
On mobile devices, higher symptoms were associated with greater engagement in activities such as entertainment, messaging, and social media, concentrated in the nighttime and occurring in fewer but longer sessions. 
On desktop, higher depressive symptoms were associated with less engagement in activities such as news, shopping, and adult content. 
Therefore, the digital profile of depression may not be merely a change in how much people do online, but in where and what they engage with. It represents a digital-context-specific pattern rather than a uniform change in overall use.

Behavioral models of depression describe these changes as a pattern of behavioral withdrawal, including reduced engagement with rewarding and valued activities, increased avoidance, and diminished contact with environmental sources of positive reinforcement ~\cite{ferster1973functional}. 
This fits the desktop results, where higher symptoms were associated with less time on news, shopping, and adult content consumption. 
Mobile devices, by contrast, are continuously at hand and support brief, on-demand use~\cite{schrock2015communicative,Jokela2015device}, remaining accessible even when a person withdraws from more structured settings. 
This may help explain why higher symptoms were associated with greater engagement with entertainment, messaging, and social media, especially in the nighttime. 
These patterns are broadly consistent with reward-based accounts of depression, in which impaired reward motivation and altered effort–cost decision-making may bias engagement toward the most readily accessible options~\cite{winer2019anhedonia,belujon2017dopamine,klein_examining_2026,treadway2012effort}, matching the concentration of associations in the continuously available mobile environment. 
Although our data did not capture the effort or reward associated with specific activities, our findings may shed light on a potential reward-based mechanism for understanding depression's environment-specific digital profile and motivate future work.

These associations were not uniform across symptom dimensions but related differently to the general, cognitive–affective, and somatic factors. 
General severity was reflected mainly in the temporal patterns of mobile use: fewer daily sessions and more nighttime activity. 
Cognitive–affective symptoms were associated with more mobile entertainment use.
The somatic dimension was primarily associated with reduced desktop engagement in activities such as news and shopping. 
These differences may help explain how different symptom dimensions relate to digital 
behavior. 

Somatic symptoms (fatigue, low energy, psychomotor slowing) correlate with reclining or 
bedbound periods~\cite{dauvilliers_hypersomnia_2013}, potentially explaining reduced desktop engagement: desktop use requires upright positioning, whereas mobile use permits 
lower-effort engagement while reclining. 
Cognitive–affective symptoms (anhedonia, worthlessness) may instead manifest as increased mobile engagement. 
Thus, dimensional decomposition suggests that device-specific patterns may reflect distinct symptom manifestations.

Our findings may explain mixed prior findings by addressing three dimensions of our study design.
First, the association runs in opposite directions across devices: greater mobile use but 
reduced desktop use, so studies that observe only a single device may reach opposite conclusions depending on which one they examine ~\cite{bettmann2021young,saeb_mobile_2015}. 
Second, studies that collapse various activities into a single screen time metric ~\cite{leaning2024smartphone} often find conflicting or null results, because the opposing associations offset one another. 
Third, self-reported use does not accurately reflect actual behavior and 
adds noise that further weakens associations ~\cite{parry_systematic_2021}. 
By observing both environments with objective traces and disaggregating by activity type and time, our design may provide a clearer view of behavioral patterns that single-device, aggregate, or self-reported measures may obscure.

These findings can inform future research and clinical practice.
For researchers, our results argue against treating screen time as a single construct: digital behavior should be measured objectively and modeled by specific devices and activities. 
In a clinical context, they highlight the value of considering an individual's digital life by inquiring about their digital habits and how and where they engage online. 
These patterns might also inform interventions: because behavioral-activation 
approaches work by gradually re-engaging people with valued activities~\cite{jacobson2001behavioral}, a potentially useful graded approach could begin from where a person already engages online and broaden toward a wider range of activities and settings, including in-person connection, to complement such strategies.

As an observational study, our design limits causal and directional inference: we cannot determine whether depressive symptoms shape web behavior, the reverse, or both.
Future experimental and intervention studies are needed to clarify these causal mechanisms.
Because depressive symptoms were largely stable within individuals over the study period, our within-person analyses were underpowered and should be treated as exploratory.
Our data came from a panel of gig workers who frequently participate in online surveys. 
To minimize potential biases associated with professional survey participation, we excluded participants showing disproportionately high survey-taking activity, and prior work indicates that web trace data from such panels reflect the most visited websites in Germany and that participants' privacy attitudes are comparable to those of non-tracked populations \cite{Kulshrestha2021WebRoutineness, Stier2020PopulistAttitudes}, supporting the representativeness of these web traces. 
The sample also consists exclusively of German adults observed over a six-month period, and its sociodemographic composition departs somewhat from the general population: gender is closely matched to the German population, but the middle ranges of age and income are overrepresented while the extremes are underrepresented~\cite{belal_examining_2026}.
Therefore, the findings may generalize less well to the youngest and oldest, and to the lowest- and highest-income individuals.
To establish the generalizability and robustness of these findings, future research should replicate this approach across diverse cultures, age groups, and clinical populations, potentially over a long time. 
We classified online activities into eight broad categories. While this captures general usage types, it overlooks specific content being consumed within webs or applications. 
Future studies should incorporate content-level classification to provide more precise distinctions.

Ultimately, our findings point beyond the specific devices studied. 
The underlying principle is general: the same condition can be expressed differently across digital contexts. 
As these environments evolve, so too is the depression profile likely to change, with each new context potentially showing its own pattern. 
The meaningful question is therefore not how much time people spend on screens, but what the depression profile looks like in each context.
This requires attention to where and what people engage with online, not merely how much.

\section{methods}

\subsection{Data collection and preprocessing}

We followed 1,490 adults over a 6-month period, combining monthly self-report surveys with continuous passively logged web traces from mobile and desktop devices. 
Participants were recruited in Germany through Bilendi (a GDPR-compliant European panel provider). 
The study protocol was reviewed and approved by the author's University Research Ethics Committee prior to data collection. 
Participants provided informed consent before participation, and we applied data minimization and anonymization procedures detailed in the ethics application. 
Each month, participants completed an online questionnaire that included the PHQ-9. 
In Wave 1, the questionnaire additionally collected demographic information, including age, gender, and income, as well as chronotype assessed using the German version of Morningness–Eveningness Questionnaire (D-MEQ)~\cite{griefahn_zur_2001}.
Each trace record included the accessed application / web domain, timestamp, and visit duration. 
The two device streams were processed and analyzed independently.

The straightlining rate in the raw data was approximately $20\%$; we adopted a conservative exclusion rule, removing 56 participants who straightlined the PHQ-9 across all 6 waves. 
The resulting PHQ-9 dataset comprised 1,434 participants and 7,390 participant-wave observations, representing 85.9\% of the 8,604 possible participant-waves, corresponding to a mean of 5.2 of 6 waves per participant (per-wave participation in \textit{SI Appendix}, Table S2). 
PHQ-9 scores had a mean of 6.18 ($SD = 5.73$), with most variance attributable to between-person 
differences (ICC$_1 = 0.794$).

For web behavior features, we extracted the 2-week window of web trace data immediately preceding the survey completion date in each PHQ-9 wave, which aligns with the 2-week timeframe specified in the survey questions. 
Sessions were defined using a 30-min inactivity threshold on both devices~\cite{Eickhoff,GoogleAnalytics2023}. 
Participant-waves with fewer than 50 usage records in the 2-week window were excluded, 
along with participants flagged as heavy survey users ($> 25\%$ completion rate). We inspected the upper tail of session durations and applied a 4-h winsorization cap: 143 of 538 desktop and 7 of 356 mobile records identified as idle artifacts were removed, and the remaining records above the threshold were capped at four hours rather than excluded to preserve plausibly real long sessions.

\begin{figure}
\centering
\includegraphics[width=\textwidth]{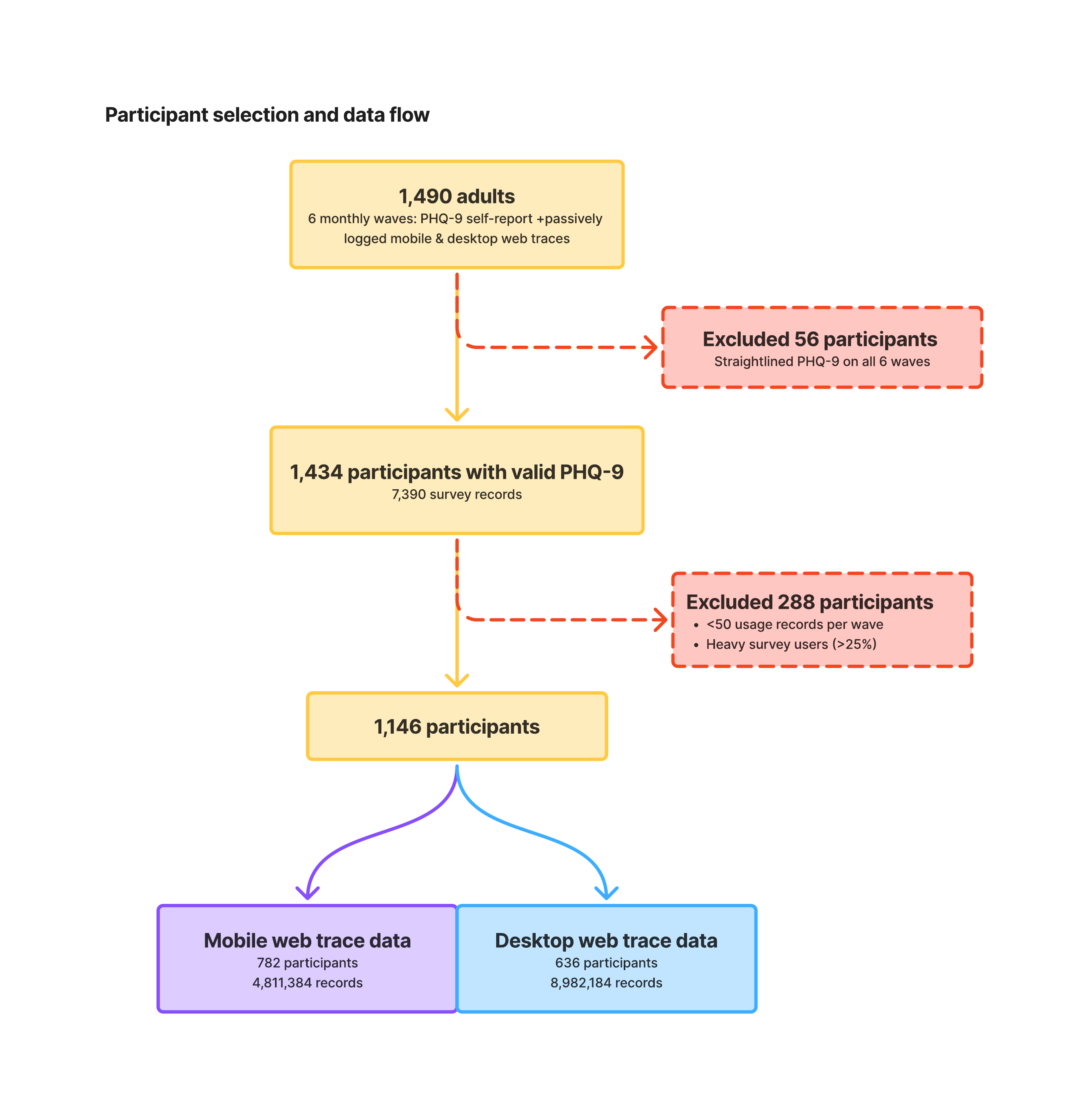}
\caption{\textbf{Participant selection and data flow.} Of 1,490 recruited adults, 56 were excluded for straightlining the PHQ-9 across all six waves, leaving 1,434 participants with valid self-report data (7,390 participant-waves). Applying the
per-device web-trace inclusion criteria then yielded the final analytic sample of 1,146 adults: participant-waves with fewer than 50 usage records were removed (653 mobile, 464 desktop), and heavy survey users ($>$25\% completion) were excluded, so that participants retaining no valid device-wave dropped out (net 288 participants excluded). The final sample comprised 782 mobile and 636 desktop contributors. Within each stream, idle-artifact records were removed (7 mobile, 143 desktop) and session durations
exceeding 4~h were winsorized (capped at 4~h and retained: 349 mobile, 395 desktop records). Full descriptive statistics appear in \emph{SI Appendix}, Tables S1--S3.}
\label{fig:si_flow}
\end{figure}

The final analytic sample comprised 1,146 adults residing in Germany, of whom 782 contributed mobile web traces (4,811,384 records) and 636 contributed desktop web traces (8,982,184 records). Of these, 272 participants contributed data from both device types (participant selection and data flow in~\ref{fig:si_flow}; full sample characteristics in \textit{SI Appendix}, Table S3).

\subsection{Bifactor model of the PHQ-9}

To examine symptom-level structure beyond the total score, we conducted confirmatory factor analysis (CFA) on the nine PHQ-9 items using the weighted least squares mean and variance adjusted (WLSMV) estimator for ordinal indicators, implemented in the \texttt{lavaan} R package~\cite{JSSv048i02}. 
Model fit was evaluated using the comparative fit index (CFI), Tucker–Lewis index (TLI), root mean 
square error of approximation (RMSEA), and standardized root mean square residual (SRMR)~\cite{hu_cutoff_1999,putnick_measurement_2016}. 
We compared unidimensional, correlated two-factor, and bifactor models using Wave 1 data (1,434 observations). 
The bifactor model demonstrated the best fit (RMSEA $= 0.051$, CFI $= 0.996$, TLI $= 0.992$, SRMR $= 0.023$) and was retained for subsequent analyses (model comparison 
in \textit{SI Appendix}, Table S4 -- S6). 
The model specifies a general depression factor loading on all nine items alongside two orthogonal specific factors: a cognitive--affective factor (anhedonia, depressed mood, 
worthlessness, concentration, and suicidal ideation) and a somatic factor (sleep, fatigue, appetite, and psychomotor change). 
Item assignments followed a German population validation of the bifactor PHQ-9~\cite{tibubos2021bidimensional}.

Full scalar measurement invariance was supported across all six waves, based on CFI and RMSEA changes from the metric to scalar model, which remained within the established measurement-invariance cutoffs ($|\Delta\text{CFI}| \leq 0.010$, $|\Delta\text{RMSEA}| \leq 
0.015$)~\cite{chen_sensitivity_2007,cheung_evaluating_2002}. 
McDonald's $\omega_{\text{hierarchical}}$ ranged from 0.77 to 0.83 across waves, estimated using the \texttt{psych} R package~\cite{zinbarg_cronbachs_2005,revelle_psych_2026}, with the general factor accounting for 81–87\% of reliable variance ($\omega_H/\omega_T$)~\cite{rodriguez_evaluating_2016}. 
Per-wave factor scores for the general, cognitive--affective, and somatic factors 
were extracted using \texttt{lavPredict}~\cite{JSSv048i02} and served as outcomes in subsequent analyses.

\subsection{Behavioral feature engineering}
To characterize the functional nature of online activity, we categorized observed web domains and mobile applications into functional categories using a two-step semi-automated procedure~\cite{belal_examining_2026,wang_lonely_2025}. 
First, domains and apps were assigned to categories by a rule-based mapping derived from the Webshrinker domain classification service~\cite{webshrinker_category_api} and major mobile app stores.
For platforms hosting multiple services, classification was performed at the sub-domain level whenever possible (e.g., \texttt{google.com} as search, \texttt{mail.google.com} as productivity), so that functionally distinct uses of the same platform were not collapsed. 
In the second step, two researchers jointly reviewed the automated labels and manually classified domains and applications that were either misclassified or left unassigned, targeting per-user coverage of at least $85\%$ of total visits. 
Inter-annotator reliability on a random subset of 200 domains yielded Cohen's $\kappa = 0.72$~\cite{landis_measurement_1977}, indicating substantial agreement. The final taxonomy classified $3,691$ web domains and $1,001$ mobile applications, covering $86\%$ of all observed visits.

For each digital device and each participant-wave, we computed three classes of features. 
Eight category-specific variables captured daily active time in each functional category (social media, messaging, entertainment, news, shopping, productivity, adult content, and gaming). 
Three session metrics captured session frequency, duration, and spacing: average daily number of sessions, average session duration, and average interval between sessions. A new session was defined after 30 minutes of inactivity, consistent with common web analytics conventions~\cite{GoogleAnalytics2023,Eickhoff}. 
Two temporal patterns captured the timing and regularity of activity: the nighttime share of activity (00:00--06:00) and interdaily stability (IS), which quantified the consistency of the daily activity pattern across days, with higher values indicating greater day-to-day regularity~\cite{luik_24-hour_2015}.

\subsection{Statistical analysis}
We examined the association between depression and online behavior using population-level linear mixed-effects models (LMMs) and group comparisons, with an exploratory within-person linear mixed-effects analysis.
 
\paragraph*{Linear mixed-effects models}
For each device and each of the three bifactor outcomes (general, cognitive--affective, somatic), we fitted a population-level LMM with the depression factor score as the outcome, behavioral features and demographic characteristics as fixed effects, and a participant-level random intercept (\texttt{lme4}~\cite{bates_fitting_2015}):

\begin{equation}
    Y_{iw} = \beta_0 + \boldsymbol{\beta}^{\mathsf{T}}\mathbf{X}_{iw}
    + \boldsymbol{\gamma}^{\mathsf{T}}\mathbf{Z}_{iw}
    + b_i + \epsilon_{iw}.
    \label{eq:lmm}
\end{equation}

where $Y_{iw}$ denotes the depression factor score for participant $i$ at wave $w$; $X_{iw}$ is the vector of behavioral features; $Z_{iw}$ is the vector of covariates; $\boldsymbol{\beta}$ and $\boldsymbol{\gamma}$ are the corresponding fixed-effect coefficients; $b_i$ is the participant-specific random intercept capturing time-invariant individual differences; and $\epsilon_{iw}$ is the residual error term.
Models were fitted by maximum likelihood using raw per-wave feature values. Covariates were gender, age group (18--30, 31--60, 60+), income group ($\leq$3000 vs.\ $>$3000 EUR), chronotype (D-MEQ: morning/neutral/evening), and survey wave. 
We report $\beta$ with 95\% CI and $p$-values, and marginal/conditional $R^2$ from the \texttt{performance} package~\cite{ludecke_performance_2021}. All variance inflation factors (VIFs) were lower than $5$ (maximum $3.4$ on mobile, $2.3$ on desktop; see \textit{SI Appendix}, Table S10 for details), indicating no problematic multicollinearity~\cite{hair_jr_multivariate_2010}. 

For the exploratory within-person analysis, behavioral features were person-mean-centered by subtracting each participant’s mean value across waves from the corresponding per-wave value. The same LMMs were then refitted using the centered behavioral features. 

\paragraph*{Group comparison}

To compare whether participants with sustained depression differed behaviorally from those with sustained minimal symptoms, we identified two extreme groups on the basis of mean PHQ-9 across the six waves using established clinical cutoffs~\cite{kroenke2001phq}: a moderate-to-severe group (mean PHQ-9 $\geq 10$; $n = 177$ mobile, $n = 125$ desktop) and a minimal-symptom group (mean PHQ-9 $\leq 5$; $n = 392$ mobile, $n = 337$ desktop). For each participant, behavioral features were averaged across waves before group comparison. For each behavioral feature, values were log-transformed using $\log(1+x)$ to reduce right skew. Interdaily stability and nighttime activity ratio were retained on their original scales. 
Group differences were tested using Welch's $t$-test, allowing for unequal variances. 
We computed Hedges' $g$ (small-sample-corrected standardized mean difference) and its 95\% bootstrap confidence interval ($5,000$ iterations). 
$P$-values were adjusted within each device’s feature set using the Benjamini--Hochberg false discovery rate procedure~\cite{benjamini_controlling_1995}.

\subsection*{Robustness checks}
Our main findings were robust across a series of checks, with only minor variation in borderline effects.
The results were stable to preprocessing choices. Varying the survey-quality threshold (20\% vs. 30\%) and the minimum-records cutoff (25 vs. 100 records per two-week window) left the primary associations unchanged. 
Alternative outlier handling approaches (winsorization and z-score removal) produced broadly consistent effect-size estimates, with mean absolute deviations in Hedges' $g$ of $0.022\text{--}0.046$ across devices. Applying a log transformation to all features produced estimates nearly identical to the primary specification (mean absolute deviation $< 0.002$). Differences in statistical significance were limited to borderline effects (original uncorrected $p = 0.03\text{--}0.06$). Agreement between the parametric and rank-based tests was $84.6\%$ for mobile and $92.3\%$ for desktop, with two and one divergent features, respectively (full results in \textit{SI Appendix}, Table S12 -- S14).

\section{Declaration of generative AI and AI-assisted technologies in the writing process}

During the preparation of this work the authors used ChatGPT for text proof reading (including spelling and grammar checks). After using this tool, the authors carefully reviewed and edited the content as necessary and take full responsibility for the final content of the published article.

\clearpage
\bibliographystyle{plain}
\bibliography{reference}

\clearpage
\begin{appendices}
\setcounter{table}{0}
\renewcommand{\thetable}{S\arabic{table}}

\begin{table}\centering
\caption{Web-trace records per wave in the final analytic sample, by device. Mobile and desktop streams were processed independently; counts reflect records retained after cleaning (participant-waves with $<$50 usage records removed; session durations winsorized at 4~h). The two streams derive from 782 (mobile) and 636 (desktop) participants, 272 of whom contributed both.}
\begin{tabular}{lrr}
\toprule
Wave & Mobile records & Desktop records\\
\midrule
1 & 1{,}016{,}934 & 1{,}928{,}398\\
2 &   830{,}343 & 1{,}659{,}486\\
3 &   824{,}164 & 1{,}136{,}573\\
4 &   770{,}116 & 1{,}377{,}379\\
5 &   675{,}914 & 1{,}607{,}203\\
6 &   693{,}913 & 1{,}273{,}145\\
\midrule
Total & 4{,}811{,}384 & 8{,}982{,}184\\
\bottomrule
\end{tabular}
\label{tab:si_webrecords}
\end{table}

\begin{table}\centering
\caption{Per-wave participation and PHQ-9 descriptive statistics. PHQ-9 total scores (possible range 0--27) by survey wave; $N$ is the number of participants completing the PHQ-9 at each wave. Across all waves there were 7{,}390 participant-wave observations (85.9\% of the 8{,}604 possible; 1{,}434 participants $\times$ 6 waves), with an overall mean of 6.18 (SD = 5.73).}
\begin{tabular}{lrrrrrr}
Wave & $N$ & Mean & SD & Median & Min & Max \\
\midrule
1 & 1{,}434 & 6.64 & 5.59 & 5 & 0 & 27 \\
2 & 1{,}300 & 6.43 & 5.86 & 5 & 0 & 27 \\
3 & 1{,}196 & 6.27 & 5.85 & 5 & 0 & 27 \\
4 & 1{,}179 & 6.08 & 5.68 & 4 & 0 & 27 \\
5 & 1{,}191 & 5.93 & 5.58 & 4 & 0 & 27 \\
6 & 1{,}090 & 5.58 & 5.62 & 4 & 0 & 27 \\
\bottomrule
\end{tabular}
\label{tab:si_phq_waves}
\end{table}

\begin{table}\centering
\caption{Sample characteristics of the analytic sample ($N=1{,}146$). Counts and column percentages.}
\begin{tabular}{lrr@{\hspace{2.5em}}lrr}
\toprule
Characteristic & $n$ & \% & Characteristic & $n$ & \% \\
\midrule
\multicolumn{3}{l}{\textit{Gender}} & \multicolumn{3}{l}{\textit{Income (monthly)}}\\
\quad Male   & 579 & 50.5 & \quad $\leq$\,€3000 & 610 & 53.2 \\
\quad Female & 567 & 49.5 & \quad $>$\,€3000    & 496 & 43.3 \\
             &     &      & \quad Missing       &  40 &  3.5 \\[2pt]
\multicolumn{3}{l}{\textit{Age}} & \multicolumn{3}{l}{\textit{Chronotype}}\\
\quad 18--30 & 101 &  8.8 & \quad Neutral & 583 & 50.9 \\
\quad 31--60 & 813 & 70.9 & \quad Morning & 441 & 38.5 \\
\quad 60+    & 232 & 20.2 & \quad Evening & 122 & 10.6 \\
\bottomrule
\end{tabular}
\label{tab:si_demographics}
\end{table}

\begin{table}\centering
\caption{Confirmatory factor analysis model comparison for the PHQ-9 (Wave 1; $N=1{,}434$). The bifactor model showed the best fit and was retained for all subsequent analyses.}
\begin{tabular}{lrrrrrr}
\toprule
Model & $\chi^2$ & df & CFI & TLI & RMSEA [90\% CI] & SRMR \\
\midrule
Unidimensional & 256.77 & 27 & 0.987 & 0.982 & 0.077 [0.069, 0.086] & 0.039 \\
Two-factor     & 236.53 & 26 & 0.988 & 0.983 & 0.075 [0.067, 0.084] & 0.038 \\
Bifactor       &  84.43 & 18 & 0.996 & 0.992 & 0.051 [0.040, 0.062] & 0.023 \\
\bottomrule
\end{tabular}
\label{tab:si_cfa}
\end{table}

\begin{table}\centering
\caption{Measurement invariance of the bifactor PHQ-9 across the six survey waves. $\Delta$CFI and $\Delta$RMSEA are relative to the preceding, less-constrained model. All $\Delta$CFI values were within the $|\Delta\text{CFI}|\le0.010$ cutoff; at the metric step RMSEA \emph{decreased} (i.e., fit improved), and the scalar step remained within both cutoffs, supporting scalar invariance.}
\begin{tabular}{lrrrrrrrr}
\toprule
Model & $\chi^2$ & df & CFI & TLI & RMSEA & SRMR & $\Delta$CFI & $\Delta$RMSEA \\
\midrule
Configural & 678.61 & 108 & 0.995 & 0.989 & 0.066 & 0.028 & --- & --- \\
Metric     & 456.25 & 183 & 0.997 & 0.997 & 0.035 & 0.032 & $+0.002$ & $-0.031$ \\
Scalar     & 662.72 & 228 & 0.996 & 0.996 & 0.039 & 0.030 & $-0.001$ & $+0.004$ \\
\bottomrule
\end{tabular}
\label{tab:si_invariance}
\end{table}

\begin{table}\centering
\caption{Reliability of the bifactor PHQ-9 by survey wave. $\omega_H$, omega hierarchical (general factor); $\omega_T$, omega total (all common factors); $\omega_H/\omega_T$, proportion of reliable variance attributable to the general factor. Estimated with the R package psych.}
\begin{tabular}{lrrr}
\toprule
Wave & $\omega_H$ & $\omega_T$ & $\omega_H/\omega_T$ \\
\midrule
1 & 0.809 & 0.943 & 0.857 \\
2 & 0.772 & 0.953 & 0.810 \\
3 & 0.815 & 0.954 & 0.854 \\
4 & 0.798 & 0.953 & 0.837 \\
5 & 0.833 & 0.953 & 0.874 \\
6 & 0.780 & 0.956 & 0.816 \\
\bottomrule
\end{tabular}
\label{tab:si_reliability}
\end{table}

\begin{table*}\centering
\caption{Population-level (raw-predictor) linear mixed models predicting the general, cognitive--affective, and somatic PHQ-9 factor scores from behavioral features and covariates, with a by-participant random intercept. $\beta$, unstandardized coefficient; $^{*}p<.05$, $^{**}p<.01$, $^{***}p<.001$. Behavioral features are listed first, covariates below. This is the primary specification.}
\footnotesize
\begin{tabular}{lrlrlrl}
\toprule
 & \multicolumn{2}{c}{General} & \multicolumn{2}{c}{Cognitive--affective} & \multicolumn{2}{c}{Somatic}\\
\cmidrule(lr){2-3}\cmidrule(lr){4-5}\cmidrule(lr){6-7}
Predictor & $\beta$ & $p$ & $\beta$ & $p$ & $\beta$ & $p$\\
\midrule
\multicolumn{7}{l}{\textit{\textbf{Mobile}}}\\
Social media & 0.013 & 0.658 & -0.014 & 0.441 & -0.010 & 0.645\\
Messaging & 0.026 & 0.552 & 0.042 & 0.156 & 0.007 & 0.848\\
Entertainment & 0.021 & 0.488 & 0.045* & 0.022 & 0.007 & 0.771\\
News & 0.021 & 0.887 & -0.041 & 0.67 & -0.127 & 0.256\\
Shopping & 0.004 & 0.974 & -0.104 & 0.227 & -0.095 & 0.343\\
Gaming & 0.011 & 0.607 & -0.011 & 0.441 & -0.023 & 0.163\\
Adult content & 0.228 & 0.315 & -0.182 & 0.193 & 0.250 & 0.121\\
Productivity & -0.083 & 0.186 & 0.026 & 0.572 & 0.018 & 0.738\\
Avg.\ session duration & -0.017 & 0.679 & -0.003 & 0.906 & 0.037 & 0.265\\
Avg.\ daily sessions & -0.020* & 0.032 & 0.003 & 0.693 & 0.005 & 0.512\\
Avg.\ time between sessions & 0.001 & 0.927 & -0.004 & 0.691 & 0.006 & 0.58\\
Nighttime usage ratio & 0.281** & 0.007 & -0.115 & 0.125 & 0.050 & 0.572\\
Interdaily stability & -0.048 & 0.717 & -0.146 & 0.123 & -0.063 & 0.572\\
\addlinespace
\multicolumn{7}{l}{\textit{Covariates}}\\
Gender (female vs male) & 0.266*** & < .001 & -0.077* & 0.013 & -0.174*** & < .001\\
Age 31--60 (vs 18--30) & -0.220. & 0.058 & -0.127* & 0.026 & -0.022 & 0.731\\
Age 60+ (vs 18--30) & -0.416** & 0.002 & -0.197** & 0.002 & -0.109 & 0.136\\
Income $>$3000 EUR (vs $\le$ 3000 EUR) & -0.182** & 0.004 & -0.073* & 0.017 & 0.070* & 0.041\\
Wave (n) & 0.004 & 0.406 & 0.011** & 0.005 & 0.001 & 0.77\\
Chronotype: morning (vs neutral) & -0.395*** & < .001 & -0.043 & 0.175 & 0.062. & 0.082\\
Chronotype: evening (vs neutral) & 0.089 & 0.394 & -0.038 & 0.452 & -0.152** & 0.008\\
\midrule
\multicolumn{7}{l}{\textit{\textbf{Desktop}}}\\
Social media & -0.003 & 0.931 & 0.024 & 0.277 & -0.012 & 0.588\\
Messaging & -0.048 & 0.636 & -0.025 & 0.693 & -0.020 & 0.768\\
Entertainment & 0.011 & 0.61 & 0.014 & 0.375 & 0.041* & 0.017\\
News & -0.039 & 0.62 & 0.001 & 0.985 & -0.121* & 0.049\\
Shopping & 0.016 & 0.736 & -0.007 & 0.843 & -0.088* & 0.035\\
Gaming & -0.038 & 0.465 & -0.036 & 0.302 & -0.015 & 0.693\\
Adult content & 0.013 & 0.681 & -0.030 & 0.204 & -0.089*** & 0.001\\
Productivity & -0.031 & 0.651 & -0.042 & 0.374 & -0.056 & 0.283\\
Avg.\ session duration & -0.020 & 0.254 & -0.003 & 0.81 & 0.001 & 0.967\\
Avg.\ daily sessions & -0.000 & 0.995 & -0.010 & 0.376 & 0.003 & 0.833\\
Avg.\ time between sessions & -0.006 & 0.437 & -0.007 & 0.28 & -0.010 & 0.163\\
Nighttime usage ratio & 0.226 & 0.126 & 0.038 & 0.732 & -0.013 & 0.917\\
Interdaily stability & 0.129 & 0.193 & 0.022 & 0.759 & -0.097 & 0.238\\
\addlinespace
\multicolumn{7}{l}{\textit{Covariates}}\\
Gender (female vs male) & 0.222*** & 0.001 & -0.102** & 0.006 & -0.149*** & < .001\\
Age 31--60 (vs 18--30) & -0.261* & 0.03 & -0.158* & 0.018 & -0.171* & 0.012\\
Age 60+ (vs 18--30) & -0.487*** & < .001 & -0.202** & 0.007 & -0.163* & 0.03\\
Income $>$3000 EUR (vs $\le$ 3000 EUR) & -0.212** & 0.002 & -0.050 & 0.174 & 0.107** & 0.004\\
Wave (n) & 0.007 & 0.203 & 0.009. & 0.07 & 0.012* & 0.027\\
Chronotype: morning (vs neutral) & -0.348*** & < .001 & -0.068. & 0.079 & 0.092* & 0.018\\
Chronotype: evening (vs neutral) & 0.023 & 0.831 & -0.082 & 0.177 & -0.049 & 0.416\\
\bottomrule
\end{tabular}
\label{tab:si_lmm_pop}
\end{table*}

\begin{table*}\centering
\caption{Within-person (person-mean-centered) linear mixed models. Behavioral features are centered on each participant's mean; covariates are as in the population-level model. $\beta$, unstandardized coefficient; $^{*}p<.05$, $^{**}p<.01$, $^{***}p<.001$. }
\footnotesize
\begin{tabular}{lrlrlrl}
\toprule
 & \multicolumn{2}{c}{General} & \multicolumn{2}{c}{Cognitive--affective} & \multicolumn{2}{c}{Somatic}\\
\cmidrule(lr){2-3}\cmidrule(lr){4-5}\cmidrule(lr){6-7}
Predictor & $\beta$ & $p$ & $\beta$ & $p$ & $\beta$ & $p$\\
\midrule
\multicolumn{7}{l}{\textit{\textbf{Mobile}}}\\
Social media & -0.012 & 0.738 & -0.032 & 0.241 & -0.012 & 0.709\\
Messaging & 0.000 & 1 & 0.013 & 0.724 & -0.022 & 0.624\\
Entertainment & -0.005 & 0.89 & 0.054* & 0.049 & -0.011 & 0.748\\
News & 0.123 & 0.463 & 0.170 & 0.19 & -0.093 & 0.546\\
Shopping & 0.011 & 0.937 & -0.116 & 0.282 & -0.061 & 0.632\\
Gaming & 0.018 & 0.477 & 0.005 & 0.801 & 0.005 & 0.831\\
Adult content & 0.736* & 0.018 & -0.411. & 0.088 & 0.297 & 0.298\\
Productivity & -0.076 & 0.242 & 0.024 & 0.635 & -0.036 & 0.542\\
Avg.\ session duration & -0.044 & 0.333 & -0.021 & 0.544 & 0.045 & 0.284\\
Avg.\ daily sessions & -0.014 & 0.211 & -0.001 & 0.874 & 0.018. & 0.076\\
Avg.\ time between sessions & 0.001 & 0.955 & -0.002 & 0.876 & 0.001 & 0.966\\
Nighttime usage ratio & 0.232* & 0.034 & -0.046 & 0.585 & 0.059 & 0.557\\
Interdaily stability & 0.027 & 0.847 & -0.117 & 0.277 & -0.028 & 0.825\\
\addlinespace
\multicolumn{7}{l}{\textit{Covariates}}\\
Gender (female vs male) & 0.259*** & < .001 & -0.074* & 0.014 & -0.181*** & < .001\\
Age 31--60 (vs 18--30) & -0.234* & 0.044 & -0.133* & 0.02 & -0.035 & 0.587\\
Age 60+ (vs 18--30) & -0.435** & 0.001 & -0.206** & 0.001 & -0.134. & 0.065\\
Income $>$3000 EUR (vs $\le$ 3000 EUR) & -0.198** & 0.002 & -0.073* & 0.016 & 0.067* & 0.049\\
Wave (time) & 0.004 & 0.399 & 0.011** & 0.005 & 0.002 & 0.637\\
Chronotype: morning (vs neutral) & -0.404*** & < .001 & -0.043 & 0.176 & 0.059. & 0.099\\
Chronotype: evening (vs neutral) & 0.104 & 0.322 & -0.049 & 0.342 & -0.150** & 0.009\\
\midrule
\multicolumn{7}{l}{\textit{\textbf{Desktop}}}\\
Social media & -0.022 & 0.633 & 0.017 & 0.642 & -0.019 & 0.66\\
Messaging & 0.084 & 0.589 & -0.076 & 0.547 & 0.016 & 0.917\\
Entertainment & 0.012 & 0.588 & 0.018 & 0.329 & 0.048* & 0.026\\
News & -0.158. & 0.081 & -0.030 & 0.686 & -0.015 & 0.867\\
Shopping & -0.009 & 0.869 & -0.025 & 0.552 & -0.104* & 0.037\\
Gaming & -0.045 & 0.472 & 0.059 & 0.252 & 0.103. & 0.091\\
Adult content & 0.029 & 0.375 & -0.027 & 0.321 & -0.100** & 0.002\\
Productivity & -0.041 & 0.615 & -0.141* & 0.032 & -0.124 & 0.11\\
Avg.\ session duration & -0.010 & 0.588 & 0.002 & 0.903 & -0.013 & 0.448\\
Avg.\ daily sessions & 0.007 & 0.688 & 0.003 & 0.86 & -0.024 & 0.177\\
Avg.\ time between sessions & -0.005 & 0.565 & -0.005 & 0.499 & -0.009 & 0.298\\
Nighttime usage ratio & 0.260 & 0.103 & -0.044 & 0.732 & -0.161 & 0.294\\
Interdaily stability & 0.165 & 0.131 & -0.070 & 0.421 & -0.096 & 0.35\\
\addlinespace
\multicolumn{7}{l}{\textit{Covariates}}\\
Gender (female vs male) & 0.221*** & 0.001 & -0.099** & 0.007 & -0.150*** & < .001\\
Age 31--60 (vs 18--30) & -0.268* & 0.025 & -0.164* & 0.014 & -0.201** & 0.003\\
Age 60+ (vs 18--30) & -0.495*** & < .001 & -0.206** & 0.005 & -0.206** & 0.006\\
Income $>$3000 EUR (vs $\le$ 3000 EUR) & -0.209** & 0.002 & -0.050 & 0.174 & 0.113** & 0.002\\
Wave (time) & 0.008 & 0.194 & 0.010* & 0.042 & 0.013* & 0.016\\
Chronotype: morning (vs neutral) & -0.346*** & < .001 & -0.071. & 0.068 & 0.081* & 0.037\\
Chronotype: evening (vs neutral) & 0.034 & 0.756 & -0.081 & 0.176 & -0.046 & 0.445\\
\bottomrule
\end{tabular}
\label{tab:si_lmm_within}
\end{table*}

\begin{table}\centering
\caption{Variance explained by the linear mixed models. $R^2_m$, marginal (fixed effects only); $R^2_c$, conditional (fixed + random). }
\begin{tabular}{lllrr}
\toprule
Device & Specification & Dimension & $R^2_m$ & $R^2_c$\\
\midrule
\multicolumn{5}{l}{\textit{\textbf{Mobile}}}\\
 & Population-level & General & 0.099 & 0.759\\
 & Population-level & Cognitive-affective & 0.027 & 0.493\\
 & Population-level & Somatic & 0.046 & 0.470\\
 & Within-person & General & 0.096 & 0.761\\
 & Within-person & Cognitive-affective & 0.021 & 0.496\\
 & Within-person & Somatic & 0.044 & 0.474\\
\midrule
\multicolumn{5}{l}{\textit{\textbf{Desktop}}}\\
 & Population-level & General & 0.098 & 0.748\\
 & Population-level & Cognitive-affective & 0.025 & 0.513\\
 & Population-level & Somatic & 0.058 & 0.420\\
 & Within-person & General & 0.097 & 0.748\\
 & Within-person & Cognitive-affective & 0.023 & 0.517\\
 & Within-person & Somatic & 0.049 & 0.422\\
\bottomrule
\end{tabular}
\label{tab:si_modelfit}
\end{table}

\begin{table}\centering
\caption{Variance inflation factors (VIF) for the behavioral features in the population-level models. All values are well below the conventional threshold of 5, indicating no problematic multicollinearity.}
\begin{tabular}{lrr}
\toprule
Feature & Mobile & Desktop\\
\midrule
Social media & 1.48 & 1.21\\
Messaging & 1.55 & 1.04\\
Entertainment & 1.34 & 1.77\\
News & 1.07 & 1.11\\
Shopping & 1.10 & 1.12\\
Gaming & 1.45 & 1.06\\
Adult content & 1.03 & 1.11\\
Productivity & 1.12 & 1.10\\
Avg.\ session duration & 3.38 & 2.34\\
Avg.\ daily sessions & 1.93 & 1.48\\
Avg.\ time between sessions & 1.91 & 1.30\\
Nighttime usage ratio & 1.30 & 1.41\\
Interdaily stability & 1.26 & 1.18\\
\bottomrule
\end{tabular}
\label{tab:si_vif}
\end{table}

\begin{table*}\centering
\caption{Group comparison of behavioral features between persistently high-depression (mean PHQ-9 $\ge$ 10) and no-depression (mean PHQ-9 $\le$ 5) participants. Features were log-transformed prior to testing to reduce skew. Means are reported on the raw (untransformed) scale for interpretability. Hedges' $g$ with 95\% bootstrap CI; $p_\text{Welch}$ from Welch's $t$-test on the log scale; $p_\text{FDR}$, Benjamini--Hochberg adjusted. Mobile: $n_\text{high}\approx177$, $n_\text{none}\approx392$; desktop: $n_\text{high}\approx125$, $n_\text{none}\approx337$ (minor variation from missing values). Positive $g$ = higher in the high-depression group.}
\footnotesize
\begin{tabular}{lrrrrrr}
\toprule
Feature & High mean & None mean & $g$ & 95\% CI & $p_\text{Welch}$ & $p_\text{FDR}$\\
\midrule
\multicolumn{7}{l}{\textit{\textbf{Mobile}}}\\
Social media (hr/day) & 0.573 & 0.424 & 0.23 & [0.05, 0.42] & 0.016 & 0.045\\
Messaging (hr/day) & 0.374 & 0.272 & 0.24 & [0.05, 0.43] & 0.017 & 0.045\\
Entertainment (hr/day) & 0.375 & 0.211 & 0.29 & [0.09, 0.48] & 0.005 & 0.023\\
News (hr/day) & 0.038 & 0.055 & -0.15 & [-0.28, 0.00] & 0.059 & 0.129\\
Shopping (hr/day) & 0.056 & 0.061 & -0.03 & [-0.18, 0.15] & 0.678 & 0.735\\
Gaming (hr/day) & 0.386 & 0.332 & 0.12 & [-0.05, 0.31] & 0.171 & 0.247\\
Adult content (hr/day) & 0.008 & 0.019 & -0.12 & [-0.23, 0.03] & 0.111 & 0.180\\
Productivity (hr/day) & 0.084 & 0.084 & 0.02 & [-0.15, 0.20] & 0.850 & 0.850\\
Avg.\ session duration (hr) & 0.909 & 0.704 & 0.37 & [0.19, 0.55] & $<$0.001 & 0.002\\
Avg.\ daily sessions (n) & 5.779 & 6.117 & -0.11 & [-0.29, 0.06] & 0.205 & 0.266\\
Avg.\ inter-session interval (hr) & 2.062 & 1.969 & 0.08 & [-0.11, 0.26] & 0.412 & 0.487\\
Nighttime usage ratio (00:00--06:00) & 0.148 & 0.110 & 0.33 & [0.15, 0.52] & $<$0.001 & 0.005\\
Interdaily stability (0--1) & 0.187 & 0.201 & -0.14 & [-0.29, 0.02] & 0.088 & 0.163\\
\midrule
\multicolumn{7}{l}{\textit{\textbf{Desktop}}}\\
Social media (hr/day) & 0.354 & 0.274 & 0.15 & [-0.05, 0.37] & 0.174 & 0.755\\
Messaging (hr/day) & 0.018 & 0.036 & -0.04 & [-0.15, 0.19] & 0.590 & 0.808\\
Entertainment (hr/day) & 0.392 & 0.326 & 0.16 & [-0.05, 0.38] & 0.143 & 0.755\\
News (hr/day) & 0.106 & 0.085 & 0.06 & [-0.17, 0.28] & 0.596 & 0.808\\
Shopping (hr/day) & 0.197 & 0.167 & 0.10 & [-0.12, 0.33] & 0.370 & 0.808\\
Gaming (hr/day) & 0.111 & 0.108 & 0.01 & [-0.20, 0.23] & 0.960 & 0.960\\
Adult content (hr/day) & 0.012 & 0.079 & -0.25 & [-0.33, -0.15] & $<$0.001 & 0.006\\
Productivity (hr/day) & 0.159 & 0.162 & -0.03 & [-0.24, 0.19] & 0.786 & 0.851\\
Avg.\ session duration (hr) & 1.259 & 1.235 & 0.07 & [-0.13, 0.27] & 0.517 & 0.808\\
Avg.\ daily sessions (n) & 2.597 & 2.683 & -0.05 & [-0.25, 0.16] & 0.635 & 0.808\\
Avg.\ inter-session interval (hr) & 2.734 & 2.779 & -0.05 & [-0.26, 0.16] & 0.629 & 0.808\\
Nighttime usage ratio (00:00--06:00) & 0.080 & 0.075 & 0.05 & [-0.17, 0.26] & 0.684 & 0.808\\
Interdaily stability (0--1) & 0.343 & 0.333 & 0.05 & [-0.14, 0.25] & 0.596 & 0.808\\
\bottomrule
\end{tabular}
\label{tab:si_groupcomp}
\end{table*}

\begin{table}\centering
\caption{Effect of alternative preprocessing on the group comparison. $r$ is the Pearson correlation of the full 13-feature Hedges' $g$ vector with the primary (log-transformed) analysis. The last column names every feature whose nominal ($p<.05$) significance changed relative to the primary analysis, its direction ($+$ higher / $-$ lower in the high-depression group), and whether it gained or lost significance. No headline feature (mobile nighttime usage, session duration, entertainment, messaging, social media; desktop adult content) changes status under any method.}
\begin{tabular}{llcl}
\toprule
Device & Preprocessing & $r$ vs.\ primary & Features changing significance\\
\midrule
\multicolumn{4}{l}{\textit{\textbf{Mobile}}}\\
 & Winsorization (1/99\%) & 0.992 & News ($-$, gained); Adult content ($-$, gained)\\
 & $z$-score outlier removal & 0.980 & \emph{none}\\
\midrule
\multicolumn{4}{l}{\textit{Desktop}}\\
 & Winsorization (1/99\%) & 0.965 & \emph{none}\\
 & $z$-score outlier removal & 0.972 & \emph{none}\\
\bottomrule
\end{tabular}
\label{tab:si_sens_summary}
\end{table}

\begin{table}\centering
\caption{Sensitivity of the one borderline feature (mobile news use) to preprocessing. Its nominal significance depends on the preprocessing choice, and it does not survive FDR correction or the rank-based test under any method; it is therefore not interpreted as a reliable association. $g$, Hedges' $g$; $p$, Welch $t$-test; $p_\text{FDR}$, Benjamini--Hochberg adjusted.}
\begin{tabular}{lrrrc}
\toprule
Preprocessing & $g$ & $p$ & $p_\text{FDR}$ & $p<.05$ (FDR)\\
\midrule
Primary (log transform) & -0.15 & 0.059 & 0.129 & no\\
Winsorization & -0.18 & 0.016 & 0.033 & yes\\
$z$-score removal & -0.16 & 0.041 & 0.076 & no\\
\bottomrule
\end{tabular}
\label{tab:si_sens_borderline}
\end{table}

\begin{table*}\centering
\caption{Rank-based robustness of the group comparison. For each feature the table reports the parametric Welch result (on log-transformed values) alongside the non-parametric Mann--Whitney $U$ test on the untransformed values. $g$, Hedges' $g$; $p_\text{FDR}$, Benjamini--Hochberg--adjusted $p$; \checkmark, significant at FDR $p<.05$. $^{\dagger}$ marks the three features where the two tests disagree: messaging and social media (mobile) and adult content (desktop) are significant parametrically but not rank-confirmed, consistent with heavy zero-inflation and right-skew that make the tied-rank test conservative. Entertainment, session duration, and nighttime usage (mobile) are confirmed by both tests.}
\footnotesize
\begin{tabular}{lrccrc}
\toprule
 & & \multicolumn{2}{c}{Welch (log)} & \multicolumn{2}{c}{Mann--Whitney}\\
\cmidrule(lr){3-4}\cmidrule(lr){5-6}
Feature & $g$ & $p_\text{FDR}$ & sig. & $p_\text{FDR}$ & sig.\\
\midrule
\multicolumn{6}{l}{\textit{\textbf{Mobile}}}\\
Social media$^{\dagger}$ & 0.23 & 0.045 & \checkmark & 0.094 & --\\
Messaging$^{\dagger}$ & 0.24 & 0.045 & \checkmark & 0.289 & --\\
Entertainment & 0.29 & 0.023 & \checkmark & 0.021 & \checkmark\\
News & -0.15 & 0.129 & -- & 0.835 & --\\
Shopping & -0.03 & 0.735 & -- & 0.835 & --\\
Gaming & 0.12 & 0.247 & -- & 0.167 & --\\
Adult content & -0.12 & 0.180 & -- & 0.963 & --\\
Productivity & 0.02 & 0.850 & -- & 0.835 & --\\
Avg.\ session duration & 0.37 & 0.002 & \checkmark & 0.001 & \checkmark\\
Avg.\ daily sessions & -0.11 & 0.266 & -- & 0.167 & --\\
Avg.\ inter-session interval & 0.08 & 0.487 & -- & 0.694 & --\\
Nighttime usage ratio & 0.33 & 0.005 & \checkmark & 0.011 & \checkmark\\
Interdaily stability & -0.14 & 0.163 & -- & 0.312 & --\\
\midrule
\multicolumn{6}{l}{\textit{\textbf{Desktop}}}\\
Social media & 0.15 & 0.755 & -- & 0.606 & --\\
Messaging & -0.04 & 0.808 & -- & 0.606 & --\\
Entertainment & 0.16 & 0.755 & -- & 0.217 & --\\
News & 0.06 & 0.808 & -- & 0.606 & --\\
Shopping & 0.10 & 0.808 & -- & 0.697 & --\\
Gaming & 0.01 & 0.960 & -- & 0.606 & --\\
Adult content$^{\dagger}$ & -0.25 & 0.006 & \checkmark & 0.606 & --\\
Productivity & -0.03 & 0.851 & -- & 0.606 & --\\
Avg.\ session duration & 0.07 & 0.808 & -- & 0.606 & --\\
Avg.\ daily sessions & -0.05 & 0.808 & -- & 0.697 & --\\
Avg.\ inter-session interval & -0.05 & 0.808 & -- & 0.606 & --\\
Nighttime usage ratio & 0.05 & 0.808 & -- & 0.697 & --\\
Interdaily stability & 0.05 & 0.808 & -- & 0.606 & --\\
\bottomrule
\end{tabular}
\label{tab:si_ranktest}
\end{table*}

\clearpage

\end{appendices}

\end{document}